\documentclass[cameraready]{Interspeech}
\usepackage{amsmath,amssymb}
\usepackage{hyperref}

\title{Emo-BPO: Emotion Bidirectional Preference Optimization for Diffusion-based Emotional TTS}

\author[affiliation={1}]{Jiacheng}{Shi}
\author[affiliation={1}]{Hongfei}{Du}
\author[affiliation={2}]{Xinyuan}{Song}
\author[affiliation={3}]{Y. Alicia}{Hong}
\author[affiliation={1}]{Yanfu}{Zhang}
\author[affiliation={1}]{Ye}{Gao}

\address{
    $^1$ College of William \& Mary, USA,
    $^2$ Emory University, USA \\
    $^3$ George Mason University, USA
}

\email{
\{jshi12, hdu02, yzhang105, ygao18\}@wm.edu,
xinyuan.song@emory.edu,
yhong22@gmu.edu
}


\usepackage{comment}

\begin{document}
\maketitle

\begin{abstract}
Diffusion-based TTS has demonstrated strong capability in emotional speech generation. However, existing emotion alignment methods primarily reinforce target emotional expressions while overlooking modeling of competing emotional modes, limiting fine-grained controllability. This issue becomes more pronounced under classifier-free guidance (CFG), which relies on contrasting conditional trajectories to shape generation. We propose Emotion Bidirectional Preference Optimization (Emo-BPO), a simple effective framework that jointly learns emotion-aligned and emotion-contrastive score functions from reordered same-text emotional pairs. Our method requires neither auxiliary reward models nor additional annotations, and can be seamlessly integrated into existing diffusion-based TTS systems to improve emotional controllability and expressiveness.
\end{abstract}

\section{Introduction}
Emotional text-to-speech synthesis (TTS) seeks to generate speech that is both intelligible and affectively expressive. Fine-grained emotional control is critical for applications such as conversational agents~\cite{voice_agent_zeng2024glm,luo2026survey}, expressive narration~\cite{zero_tts_wang2023neural, chen2026polybench}, and human machine interaction~\cite{li2025umetts, sun2026activation}. Achieving this requires modeling stress, intonation, rhythm, and their interactions within an emotional space, rendering emotional TTS a multi-dimensional generation problem.

Early neural TTS architectures established the foundations of high-quality synthesis. Tacotron~\cite{li2024mm_tacotron} introduced end-to-end sequence-to-sequence acoustic modeling, FastSpeech~\cite{diatlova2023emospeech} improved efficiency via non-autoregressive duration prediction, and VITS~\cite{zhao2023vits} unified acoustic and waveform modeling in a variational framework. Emotional control in these systems is achieved through categorical embeddings or conditioning vectors, which limits fine-grained expressiveness and structured emotion manipulation. Recent advances adopt large language model (LLM)-based architectures and diffusion modeling to improve controllability and generalization. LLM-based TTS~\cite{du2024cosyvoice1,du2024cosyvoice,wang2025spark,guo2024fireredtts,lin2026you} demonstrates strong zero-shot capability and synthesis quality. Diffusion approaches, such as EmoMix~\cite{tang2023emomix} and EmoDiff~\cite{guo2023emodiff}, incorporate emotion conditioning into the denoising process for controllable generation. However, these methods rely on predefined labels or heuristic conditioning and do not optimize human perceptual preference.

Preference-based optimization reframes emotional alignment as a learning problem defined by human judgments. Emo-DPO~\cite{gao2025emo} performs direct preference optimization on paired samples without reward modeling. EMORL-TTS~\cite{li2025emorl_emorl-tts} formulates emotional alignment as reinforcement learning with composite reward functions. EASPO~\cite{shi2025emotion-easpo} incorporates stepwise preference signals to guide the diffusion denoising process. Despite these advances, a structural limitation persists in diffusion-based preference alignment. Diffusion models commonly rely on classifier-free guidance (CFG)~\cite{ho2022classifier_CFG}, which combines conditional and unconditional predictions at each denoising step. While existing methods optimize the conditional trajectory toward preferred outputs, they do not explicitly model divergence from non-preferred emotional modes. This asymmetric optimization underutilizes the contrastive mechanism inherent in CFG and may weaken inter-emotion discrimination. This raises a central question:
\textit{Can diffusion-based emotional TTS benefit from explicitly modeling both preference-aligned and preference-contrastive trajectories?}

We propose Emotion Bidirectional Preference Optimization (Emo-BPO), a structured framework for diffusion-based emotional TTS. Emo-BPO constructs reordered same-text emotional pairs to jointly learn emotion-aligned and emotion-contrastive score functions. During inference, these branches are integrated through a contrastive guidance formulation compatible with CFG, strengthening target emotional trajectories while suppressing competing modes. Emo-BPO does not require auxiliary reward models or additional annotations and can be seamlessly integrated into existing diffusion-based TTS systems. \textbf{Our contributions are as follows:}
(1) We identify a structural asymmetry in existing diffusion-based preference alignment methods, namely the absence of explicit modeling of emotion-contrastive trajectories under classifier-free guidance.
(2) We propose Emo-BPO, a bidirectional diffusion framework that jointly learns emotion-aligned and emotion-contrastive score functions from reordered same-text pairs, enabling improved emotional controllability without auxiliary reward models or additional datasets.
\section{Related Work}

\textbf{Emotion Alignment in Diffusion-based TTS.}
Prior emotion-controllable TTS methods rely on explicit conditioning or structured emotion representations. EmoSpeech~\cite{diatlova2023emospeech} augments FastSpeech2 with emotion embeddings and conditional attention for token-level intensity modeling. EmoSphere++\cite{cho2025emosphere++} introduces a VAD-based spherical emotion vector for controllable zero-shot style modulation. EmoVoice\cite{yang2025emovoice} adopts an LLM-based framework for natural language emotion prompting with phoneme-token co-generation. CosyVoice~\cite{du2024cosyvoice1} proposes an instruction-conditioned LLM-TTS model for controllable zero-shot synthesis, while CosyVoice2~\cite{du2024cosyvoice} scales this approach through large-scale instruction tuning to unify controllability and generalization in a single autoregressive system. Recent studies explore preference-driven optimization to better align synthesized speech with human perception. Emo-DPO~\cite{gao2025emo} applies direct preference optimization on paired samples, EMORL~\cite{li2025emorl_emorl-tts} employs reinforcement learning with multi-objective rewards, and EASPO~\cite{shi2025emotion-easpo} leverages stepwise preference signals for diffusion-based generation. However, these methods reinforce target emotions without modeling separation from competing emotional modes, limiting controllability in diffusion processes where multiple emotional trajectories coexist. We address this limitation with Emo-BPO, which learns complementary emotion-aligned and emotion-contrastive score functions from reordered same-text pairs without auxiliary reward models.
\begin{table*}
\vspace{-6mm}
\centering
\caption{Objective evaluation: Comparison of Emo-BPO with representative emotional TTS baselines in terms of emotion similarity (Emo SIM), prosody similarity (Prosody SIM), intelligibility, naturalness (UTMOS), and speech emotion recognition accuracy. Emo-BPO achieves the best performance on Emo SIM, Prosody SIM, UTMOS, and mean emotion recognition accuracy while maintaining competitive intelligibility. Results are averaged over three independent runs.}
\resizebox{0.98\textwidth}{!}{
\begin{tabular}{lcccc|cccccc}
\toprule
\textbf{Models} 
& \textbf{Emo SIM} 
& \textbf{Prosody SIM} 
& \textbf{Intelligibility} 
& \textbf{UTMOS} 
& \multicolumn{6}{c}{\textbf{Speech Emotion Recognition}} \\

\cmidrule(lr){6-11}
& & & & 
& Neutral 
& Angry 
& Happy 
& Sad 
& Surprise 
& Mean \\

\midrule
emospeech~\cite{diatlova2023emospeech} 
& 96.35 & 3.39  & 7.13  & 4.24
& 0.71& 0.64& 0.58& 0.60& 0.52& 0.61 \\

cosyvoice~\cite{du2024cosyvoice1} 
& 97.07  & 3.64 & 4.32  & 4.41 
& 0.69 & 0.76 & 0.61 & 0.74 & 0.65 & 0.69 \\

cosyvoice2~\cite{du2024cosyvoice} 
& 98.47 & 3.78 & \textbf{3.77} & 4.43 
& 0.88& 0.87& 0.72& 0.83& 0.78 & 0.82 \\

Emosphere++~\cite{cho2025emosphere++} 
& 98.16 & 3.69 & 4.94 & 4.35 
& 0.79& 0.81 &0.67 &0.77 &0.69 & 0.75 \\

EmoVoice~\cite{yang2025emovoice} 
& 98.59& 3.67& 4.16& 4.39 
& 0.87& 0.85 &0.76 &0.80 &0.79 & 0.81 \\
\midrule
Emo-BPO 
& \textbf{99.23} 
& \textbf{3.85} 
& 3.84 
& \textbf{4.52} 
& \textbf{0.91} 
& \textbf{0.89} 
& \textbf{0.82} 
& \textbf{0.87} 
& \textbf{0.84} 
& \textbf{0.87} \\

\bottomrule
\end{tabular}
}
\label{objective}
\vspace{-0.2cm}
\end{table*}

\begin{table*}[t]
\centering
\begin{minipage}[t]{0.62\textwidth}
\vspace{0pt}
\centering
\resizebox{1\textwidth}{!}{
\begin{tabular}{lccc|cccccc}
\toprule
\textbf{Models} 
& \textbf{MOS} 
& \textbf{Emo MOS} 
& \textbf{MOS EC} 
& \multicolumn{6}{c}{\textbf{Speech Emotion Recognition}} \\

\cmidrule(lr){5-10}
& & & 
& Neutral 
& Angry 
& Happy 
& Sad 
& Surprise 
& Mean \\

\midrule
emospeech~\cite{diatlova2023emospeech} 
& 2.55 & 2.79 & 2.83
& 0.66 & 0.34 & 0.25 & 0.58 & 0.57 & 0.47 \\

cosyvoice2~\cite{du2024cosyvoice} 
& 3.61 & 3.93 & 3.75
& 0.81 & 0.58 & 0.63 & 0.71 & 0.73 & 0.68 \\

Emosphere++~\cite{cho2025emosphere++} 
& 3.46 & 3.47 & 3.28
& 0.69 & 0.78 & 0.58 & 0.72 & 0.58 & 0.66 \\

EmoVoice~\cite{yang2025emovoice} 
& 3.39 & 3.64 & 3.42
& 0.71 & 0.57 & 0.55 & 0.59 & 0.65 & 0.61 \\

\midrule
Emo-BPO 
& \textbf{3.93} 
& \textbf{4.25} 
& \textbf{3.92} 
& \textbf{0.86} 
& \textbf{0.78} 
& \textbf{0.70} 
& \textbf{0.78} 
& \textbf{0.85} 
& \textbf{0.79} \\

\bottomrule
\end{tabular}
}
\caption{Subjective evaluation: Emo-BPO outperforms representative baselines in MOS, Emo MOS, MOS EC, and mean human-recognized emotion accuracy.}
\label{tab:subjective}
\end{minipage}
\hfill
\begin{minipage}[t]{0.375\textwidth}
\vspace{0pt}
\centering
\resizebox{1\textwidth}{!}{
\begin{tabular}{l c c c c}
\toprule
\textbf{Model} & \textbf{Emo SIM} & \textbf{Pro. SIM} & \textbf{WER} & \textbf{UTMOS} \\
\midrule
\textbf{Emo-BPO} & \textbf{99.23} 
& \textbf{3.85} 
& \textbf{3.84}
& \textbf{4.52} \\
Grad-TTS w/ finetune  &  
96.68 &
3.42 &
6.71 &
4.18
  \\
\midrule
\quad w/out Contrastive Branch         &  
97.41 &
3.67 &
4.48 &
4.32
  \\
\quad w/out Late-Step Schedule    &  
98.76 & 
3.82 & 
4.29 & 
4.45
  \\
\bottomrule
\end{tabular}
}
\caption{Ablation study: Removing the contrastive branch or scheduling strategy consistently degrades performance across all metrics, validating the essential contributions of both components.
}
\label{tab:ablation_mian}
\end{minipage}
\vspace{-4mm}
\end{table*}

\section{Proposed Method}
\begin{figure}[!t]
  \centering
  \includegraphics[scale=0.48]{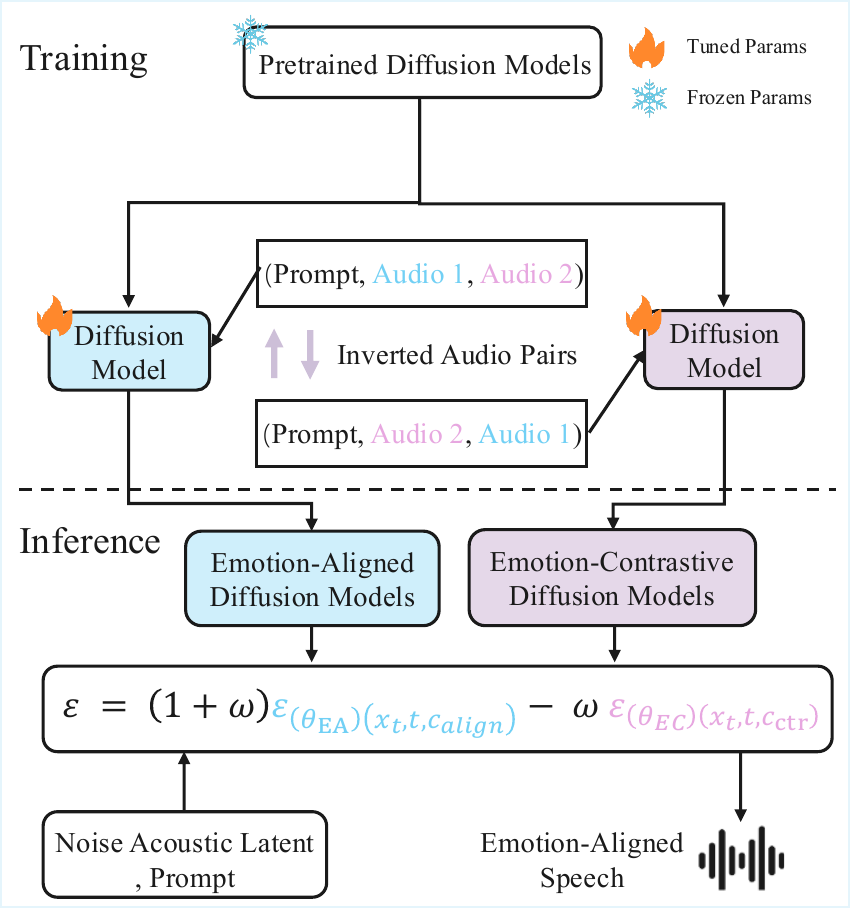}
  \caption{Overview of Emo-BPO. Training constructs inverted same-text emotional pairs to jointly learn emotion-aligned (EA) and emotion-contrastive (EC) diffusion branches while partially fine-tuning a pretrained backbone. At inference, EA and EC predictions are combined via contrastive guidance under classifier-free guidance, reinforcing target emotional trajectories and suppressing competing emotional modes.
  }
  \label{fig:pipline}
  \vspace{-4mm}
\end{figure}
\subsection{Preliminaries: Classifier-Free Guidance}

To ground our formulation in a principled probabilistic framework, we briefly revisit classifier-free guidance (CFG) from the perspective of score-based diffusion models. Consider a conditional diffusion model parameterized by an epsilon prediction network $\epsilon_\theta(x_t, c, t)$, where $x_t$ denotes the noisy acoustic representation at timestep $t$ and $c$ denotes the conditioning variable (e.g., text, speaker identity, and emotion). Under the standard variance-preserving diffusion formulation, the predicted noise is proportional to the conditional score function:
\begin{equation}
\nabla_{x_t} \log p_\theta(x_t \mid c)
=
-
\frac{1}{\sigma_t}
\epsilon_\theta(x_t, c, t).
\end{equation}

Thus, denoising can be interpreted as iteratively transporting samples along the conditional score field, moving them toward regions of higher probability under $p(x \mid c)$. Classifier-free guidance modifies this score field by introducing a contrastive correction between two conditions. Specifically, instead of relying solely on $\nabla_{x_t} \log p_\theta(x_t \mid c)$, CFG replaces it with
\begin{equation}
\nabla_{x_t} \log p_\theta(x_t \mid c)
+
\omega
\nabla_{x_t}
\log
\frac{
p_\theta(x_t \mid c)
}{
p_\theta(x_t \mid c')
},
\end{equation}
where $c’$ denotes a reference (e.g., unconditional) condition and $\omega \ge 0$ controls the guidance strength. This expression reveals that CFG performs gradient ascent on a log-likelihood ratio, increasing the probability under target condition $c$ and decreasing it under reference condition $c’$. In epsilon parameterization, the above contrastive score formulation yields
\begin{equation}
\epsilon_\theta^\omega(x_t, t)
=
(1+\omega)\epsilon_\theta(x_t,c,t)
-
\omega \epsilon_\theta(x_t,c',t),
\end{equation}
which induces a directional shift of the denoising trajectory in acoustic space. This log-ratio view of CFG provides a natural basis for emotion control. In conventional CFG, the reference condition $c’$ is typically unconditional. However, emotional speech synthesis often requires stronger separation between competing prosodic modes. This motivates our bidirectional preference formulation, where the reference condition is replaced with an emotion-contrastive condition to explicitly model inter-emotion gradients.
\subsection{Emotion Bidirectional Preference Optimization}
Existing preference-based emotional TTS methods typically optimize only the emotion-aligned direction, increasing likelihood under the target emotion without explicitly separating competing emotional modes, which limits controllability. Motivated by the log-likelihood ratio view of classifier-free guidance, we propose Emo-BPO, a contrastive diffusion framework that models directional preference between emotional conditions. Emo-BPO jointly learns (i) an emotion-aligned objective that increases likelihood under the desired emotion and (ii) an emotion-contrastive objective that suppresses alternative emotions. Intuitively, the aligned branch forms an attractive energy basin, while the contrastive branch induces repulsive gradients, enlarging inter-emotion margins. The bidirectional formulation is realized through preference-based optimization during training and contrastive guidance at inference.

\subsubsection{Train with Emo-BPO}

An important observation of Emo-BPO is that bidirectional emotional supervision can be realized without introducing additional annotations or auxiliary objectives. Given standard emotion-labeled corpora, complementary training signals naturally arise by reordering audio samples that share identical linguistic content but differ in emotion. Specifically, for a text condition $c$, we construct pairs of speech samples $(a_1, a_2)$ with identical text but different emotion labels. Without loss of generality, let $a_1$ express the target emotion and $a_2$ express an alternative emotion. The ordered tuple $(c, a_1, a_2)$ defines the emotion-aligned direction, while reversing the order to $(c, a_2, a_1)$ produces the complementary contrastive direction. This reordering provides supervision for both emotional attraction and emotional separation under matched linguistic conditions. To capture these two opposing directions, we optimize two dedicated diffusion branches. The emotion-aligned branch models the conditional score field associated with the target emotion. For a clean audio sample $a_1$, we follow the standard forward diffusion process to obtain its noisy version $x_t$, and minimize the denoising objective
\begin{equation}
\mathcal{L}_{\mathrm{align}}
=
\mathbb{E}_{t,\epsilon}
\Big[
\lVert
\epsilon -
\epsilon_{\theta_{\mathrm{EA}}}(x_t, c_{\mathrm{align}}, t)
\rVert_2^2
\Big].
\end{equation}
where $\epsilon$ denotes Gaussian noise, $t$ is the diffusion timestep, $c_{\mathrm{align}}$ represents the target-emotion condition, and $\epsilon_{\theta_{\mathrm{EA}}}$ is the score predictor of the aligned branch. Symmetrically, the emotion-contrastive branch models score directions associated with competing emotional characteristics. Using the alternative-emotion sample $a_2$, we minimize
\begin{equation}
\mathcal{L}_{\mathrm{ctr}}
=
\mathbb{E}_{t,\epsilon}
\Big[
\lVert
\epsilon -
\epsilon_{\theta_{\mathrm{EC}}}(x_t, c_{\mathrm{ctr}}, t)
\rVert_2^2
\Big].
\end{equation}
where $c_{\mathrm{ctr}}$ denotes the alternative-emotion condition and $\epsilon_{\theta_{\mathrm{EC}}}$ is the corresponding score predictor. Although both branches share the same diffusion architecture, they correspond to distinct optimization objectives: the aligned branch increases likelihood under the target emotion, while the contrastive branch models separation from competing emotional modes. Optimizing a single parameterization to satisfy both directions may result in a compromised update that weakens emotional controllability. Emo-BPO therefore maintains separate parameterizations for the aligned and contrastive directions, enabling clearer inter-emotion gradient estimation exploited during inference.
\subsubsection{Inference with Emo-BPO}

During inference, the emotion-aligned and emotion-contrastive score predictors are combined via contrastive guidance to form the final noise estimate. At diffusion timestep $t$, we compute
\begin{equation}
\epsilon^{\omega}(x_t, t)
=
(1+\omega(t))\,
\epsilon_{\theta_{\mathrm{EA}}}(x_t, c_{\mathrm{align}}, t)
-
\omega(t)\,
\epsilon_{\theta_{\mathrm{EC}}}(x_t, c_{\mathrm{ctr}}, t),
\end{equation}
where $\epsilon_{\theta_{\mathrm{EA}}}$ and $\epsilon_{\theta_{\mathrm{EC}}}$ denote the aligned and contrastive branches, respectively. The subtraction term suppresses competing emotional characteristics, while the amplified aligned term strengthens the target emotion, resulting in a bidirectional emotional steering effect. To progressively refine emotional expression during denoising, we adopt a late-step guidance schedule defined as $\omega(t)=1-\frac{t}{T}$, where $T$ denotes the total number of diffusion steps. This schedule applies weaker emotional guidance at early timesteps to preserve coarse acoustic structure and gradually increases guidance strength at later timesteps to enhance fine-grained prosodic and emotional cues. Such late-stage amplification stabilizes generation while improving emotional controllability.

\section{Experiments}
\subsection{Datasets and Experimental Setup}
We train our model on two English emotional speech corpora: the Emotional Speech Database (ESD)~\cite{zhou2022emotional_ESD_dataset} and EmoVoiceDB~\cite{yang2025emovoice}. EmoVoiceDB contains approximately 22,100 emotion-labeled utterances (about 40 hours of speech) from multiple speakers, covering diverse expressive patterns and prosodic variations. We adopt the official train/test split provided by the dataset. The English subset of ESD comprises 10 speakers expressing five emotions (Angry, Happy, Sad, Surprise, and Neutral), with 350 utterances per speaker per emotion, totaling 17,500 utterances (approximately 12 hours of speech). We follow the official 8:1:1 train/validation/test split protocol per speaker–emotion~\cite{shi2025clep, xiao2025listen}. We construct preference pairs using utterances with identical text content, following the text-matched pairing strategy of Emo-DPO~\cite{gao2025emo}. For each text prompt, audio samples expressing the target emotion are treated as preferred, while samples with other emotion labels are treated as dis-preferred.

Our backbone TTS system is Grad-TTS~\cite{popov2021grad} operating on 80-dimensional mel-spectrograms. We freeze the text encoder and duration predictor and update only the diffusion decoder (score network). The model is initialized from the Grad-TTS pre-training configuration and optimized using Adam with a learning rate of $1\times10^{-4}$ and a batch size of 16. Training is performed on randomly cropped 2-second mel-spectrogram segments~\cite{xiao2025xlsr}. Waveforms are synthesized
using a pretrained HiFi-GAN vocoder~\cite{yang2023hifi-codec}. During inference, we employ a late-step guidance coefficient defined as $\omega(t)=1-\frac{t}{T}$, where $T$ denotes the total number of diffusion steps. This schedule applies weaker guidance at early denoising stages to preserve coarse acoustic structure and gradually increases the guidance strength at later stages to refine fine-grained emotional prosody.

\begin{figure}[t]
\centering
\begin{minipage}{\linewidth}
\centering
\includegraphics[width=0.85\linewidth]{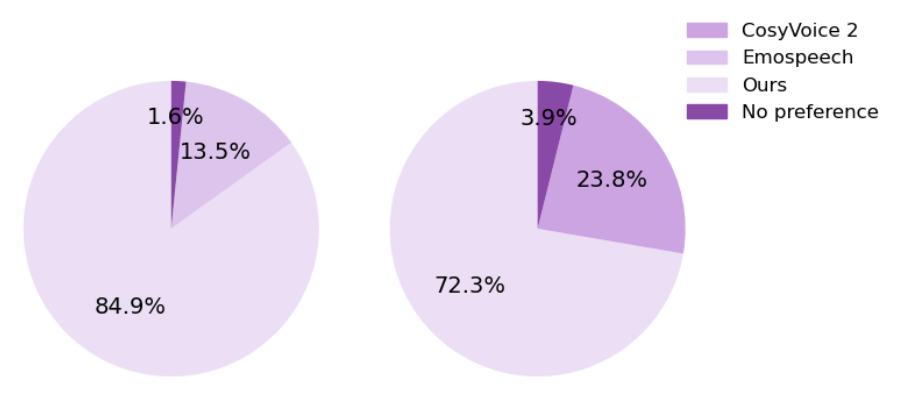}
  \caption{Comparison of subjective evaluation results from AB preference tests: 1) left: Emospeech vs. Emo-BPO and 2) right: CosyVoice2 vs. Emo-BPO
  }
\label{fig:sub_phi}
\end{minipage}


\begin{minipage}{\linewidth}
\centering
\includegraphics[width=0.8\linewidth]{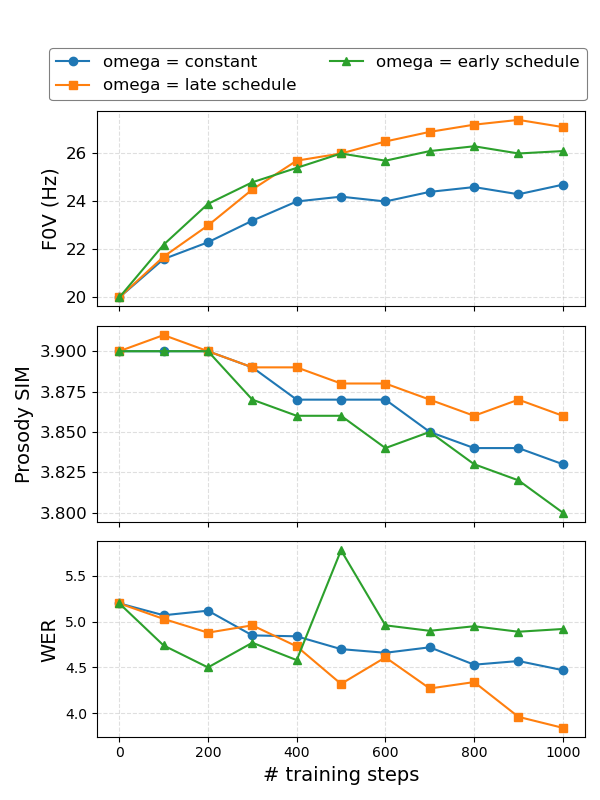}
  \caption{ Ablation on guidance scheduling. Late scheduling yields higher F0V, better prosody similarity, and lower WER, while early scheduling causes prosody degradation and transient instability.
  }
\label{fig:guidance}
\vspace{-6mm}
\end{minipage}
\end{figure}

\subsection{Evaluation Metrics}

We evaluate the proposed system using both objective and subjective metrics to assess intelligibility, prosodic consistency, emotional accuracy, and perceptual quality.

\noindent\textbf{Objective Evaluation.}
Speech intelligibility is measured by applying Whisper-Large-v3 to transcribe generated audio and computing word error rate (WER) against reference transcripts. Prosodic similarity is quantified using AutoPCP~\cite{autopcp_barrault2023seamless}, which estimates utterance-level similarity between generated and ground-truth speech; scores are averaged over the test set. Emotional similarity is evaluated by extracting embeddings with emotion2vec-base~\cite{ma2024emotion2vec} from generated and reference speech, computing cosine similarity, and reporting the mean value (EMO SIM). For automatic emotion recognition~\cite{lin2025emo,lin2025mitigating}, a pretrained speech emotion classifier predicts labels from generated utterances; accuracy is computed by averaging correctness across samples. Perceptual naturalness is further assessed using UTMOS~\cite{saeki2022utmos}, an automatic mean opinion score estimator.

\noindent\textbf{Subjective Evaluation.}
Subjective experiments involve 30 participants and include Mean Opinion Score (MOS), Emotion MOS, Emotional consistency, AB preference testing, and an Emotion Accuracy Test (EAT-EMO). In MOS, listeners rate overall speech naturalness on a 5-point scale (1: poor, 5: excellent). Emotion MOS measures the perceived similarity between generated and target emotions on the same scale. MOS EC measures consistency between the generated audio
and the instruction (emotion + text). In AB tests, listeners select the preferred sample between two systems based on quality and emotional expressiveness, with balanced emotion categories across comparisons. In Emotion Accuracy Test participants identify the perceived emotion of each sample, and accuracy is computed against the target labels. All subjective evaluations use emotion-balanced samples.

\noindent\textbf{Results and Discussion.}
Tabs.~\ref{objective} and \ref{tab:subjective} report objective and subjective results. Emo-BPO achieves the best performance in emotion similarity, prosody similarity, UTMOS, and mean emotion recognition accuracy, indicating stronger emotional alignment and perceptual quality. Fig.~\ref{fig:sub_phi} further shows clear listener preference over EmoSpeech and CosyVoice2. These gains are consistent with the bidirectional score shaping of Emo-BPO, which enlarges inter-emotion margins through complementary attractive and repulsive gradients while preserving naturalness. Although CosyVoice2 attains the lowest WER due to its autoregressive LLM-based design prioritizing linguistic fidelity, Emo-BPO maintains competitive intelligibility and substantially improves emotional expressiveness. Overall, combining emotional attraction and separation with progressive guidance scheduling enables more controllable emotional TTS. Audio samples are available online.\footnote{\url{ https://jiachengqaq.github.io/emo-bpo/}}

\subsection{Ablation Study.} 
\textbf{Ablation on Model Components.}
As shown in Tab.~\ref{tab:ablation_mian}. We conduct an ablation study to evaluate the roles of the contrastive branch and late-step scheduling. Removing the contrastive branch reduces emotion similarity, indicating weaker inter-emotion discrimination. Eliminating the scheduling mechanism degrades prosody similarity and increases WER, reflecting unstable guidance during denoising. Overall, both components are essential for achieving strong emotional alignment while preserving speech quality.

\noindent\textbf{Ablation on Guidance Scheduling.}
As shown in Fig.~\ref{fig:guidance}.We compare constant, early, and late guidance schedules in terms of emotional strength (F0V), prosody similarity, and intelligibility.  Late scheduling achieves the highest F0V at convergence while maintaining higher prosody similarity and the lowest WER. In contrast, early scheduling, defined as 
$\omega(t)=\frac{t}{T}$, increases F0V rapidly at early stages 
but leads to prosody degradation and a transient WER spike, 
indicating structural instability under strong guidance in 
high-noise denoising. Overall, progressively strengthening guidance at later stages better balances emotional expressiveness and speech fidelity.

\section{Conclusion}
We introduced Emo-BPO, a diffusion-based framework for emotional TTS that addresses a key limitation of existing preference-driven approaches: they reinforce target emotions but do not explicitly model separation from competing emotional tendencies under classifier-free guidance. Emo-BPO learns complementary emotion-aligned and emotion-contrastive score functions from reordered same-text pairs and combines them with progressive guidance scheduling during inference. Experiments demonstrate improvements in emotional alignment, prosodic control, and perceptual quality while maintaining competitive intelligibility. These results indicate that jointly modeling emotional attraction and separation leads to controllable and expressive emotional speech generation.

\section{Generative AI Use Disclosure}
The authors used generative AI tools only for grammar correction, polishing, and readability improvement. These tools were not used to generate research ideas, methodology, experimental results, analyses, or conclusions, nor to produce a significant portion of the manuscript. All authors reviewed and approved the final manuscript and take full responsibility for its content.

\bibliographystyle{IEEEtran}
\bibliography{mybib}

\end{document}